\def\CLASSINPUTbaselinestretch{0.93}
\documentclass[conference]{IEEEtran}
\IEEEoverridecommandlockouts

\usepackage{cite}
\usepackage{amsmath,amssymb,amsfonts}
\usepackage{algorithmic}
\usepackage{graphicx}
\usepackage{textcomp}
\usepackage{iftex}
\usepackage{placeins}
\usepackage{comment}
\usepackage{stfloats}
\usepackage[unicode,hidelinks,bookmarks=false]{hyperref}
\ifPDFTeX
\else
    \usepackage{fontspec}
\fi

\newcommand{\setcompactspacing}{%
    \setlength{\abovedisplayskip}{0.85ex plus 2pt minus 1pt}%
    \setlength{\belowdisplayskip}{0.85ex plus 2pt minus 1pt}%
    \setlength{\abovedisplayshortskip}{0.4ex plus 1.5pt minus 0.5pt}%
    \setlength{\belowdisplayshortskip}{0.6ex plus 1.5pt minus 0.5pt}%
    \setlength{\jot}{2pt}%
}
\AtBeginDocument{\setcompactspacing}

\def\BibTeX{{\rm B\kern-.05em{\sc i\kern-.025em b}\kern-.08em
    T\kern-.1667em\lower.7ex\hbox{E}\kern-.125emX}}
\begin{document}

\title{Online Material-Labeled Environment\\
Reconstruction via Bayesian Multipath Attribution\\
for Low-Altitude ISAC
}

\author{\IEEEauthorblockN{Meihui Liu\textsuperscript{1},
Shu Sun\textsuperscript{1}, Ruifeng Gao\textsuperscript{2}, Qiuming Zhu\textsuperscript{3}}
\IEEEauthorblockA{\footnotesize\linespread{0.92}\selectfont
\textsuperscript{1}School of Information Science and Electronic Engineering, Shanghai Jiao Tong University, Shanghai, China\\
\textsuperscript{2}School of Transportation and Civil Engineering, Nantong University, Nantong, China\\
\textsuperscript{3}College of Electronic and Information Engineering, Nanjing University of Aeronautics and Astronautics, Nanjing, China\\
Corresponding author: Shu Sun (Email: shusun@sjtu.edu.cn)\vspace{-0.6em}}
}

\maketitle

\begin{abstract}

Environment reconstruction for low-altitude integrated sensing and
communications (ISAC) has largely focused on geometry-centric maps,
overlooking material-dependent propagation effects. Material-labeled
reconstruction is therefore a key step toward propagation-aware mapping,
enabling more physically grounded channel prediction and uncrewed aerial
vehicle (UAV) networking.  However, constructing such maps from wireless multipath observations is challenging in outdoor multi-building scenarios because multipath components (MPCs) from different facades are mixed, path-to-facade attribution is uncertain, and UAV measurements arrive sequentially under time-varying observation geometries.
To address these challenges, we propose a unified online probabilistic framework that represents each reflecting facade as a virtual anchor (VA) and couples Bayesian VA localization, multipath attribution, and material inference. The Bayesian front end estimates facade-level geometry and computes soft MPC-to-VA attribution probabilities using a specular-diffuse likelihood model, thereby accounting for both dominant specular paths and diffuse surface-interacted components. These attribution probabilities are used to construct attribution-aware MPC representations, which are aggregated in a VA-centric manner and mapped by a material inference network to facade-level material evidence. The resulting evidence is recursively fused through an online Bayesian update to produce stable material posteriors and material-labeled environment maps. Ray-tracing simulations in a representative urban street scenario show that the proposed method {substantially} outperforms a no-attribution baseline, achieves 93.75\% final facade-level material accuracy on a held-out UAV trajectory, and maintains accurate VA-based facade localization.

\end{abstract}

\begin{IEEEkeywords}
Low-altitude ISAC, material-labeled environment reconstruction, Bayesian multipath attribution, virtual-anchor localization, propagation-aware mapping.
\end{IEEEkeywords}

\vspace{8pt}
\section{Introduction}

The rapid growth of the low-altitude economy is accelerating the deployment of uncrewed aerial vehicles (UAVs), urban air mobility platforms, and other aerial intelligent systems \cite{jiang2025isac_low_altitude_economy}. These platforms require wireless networks that provide connectivity while continuously sensing the propagation environment for adaptive decision making. Integrated sensing and communications (ISAC), including cooperative sensing architectures, offers a natural framework for this purpose by reusing wireless infrastructure for both communication and radio sensing \cite{LIU2026130}. A central task is to construct propagation-aware maps that capture the structures and propagation-relevant properties shaping low-altitude wireless channels.

In building-dense low-altitude corridors, radio propagation is strongly shaped by non-line-of-sight (NLoS) multipath components (MPCs) from building facades, ground surfaces, and other urban structures. Rather than {merely acting as channel impairments}, MPCs carry both geometric and material-sensitive cues. Specifically, delay, angle, and motion consistency constrain the locations and orientations of interacting surfaces \cite{leitinger2019belief_propagation_slam,sun2026modeling_land_to_ship}, whereas path gain, scattering behavior, and attenuation reflect material-dependent electromagnetic responses \cite{itu2025p2040,zhang2026diffuse_scattering_measurements}. Thus, MPCs provide a communication-native basis for material-labeled environment maps that support channel prediction, blockage reasoning, beam management, and radio digital twins.

Despite this potential, material-aware environment reconstruction from wireless MPCs remains underexplored. Existing geometry-centric methods, including geometric inversion \cite{chang2025environment_reconstruction_thz}, {virtual-anchor (VA)} mapping \cite{leitinger2019belief_propagation_slam,liu2026fusion_monostatic_bistatic_isac}, and learning-based scatterer reconstruction \cite{song2026dynamic_environment_reconstruction_isac,yang2025envrecongpt}, mainly infer reflection points, structural map features, or sparse scattering representations from channel observations, but do not assign material-dependent propagation evidence to reconstructed surfaces.
A recent terahertz monostatic sensing study \cite{lyu2025hybrid_channel_modeling_thz} demonstrated that wireless MPCs can support joint geometric reconstruction and material identification under controlled indoor sensing conditions. However, unlike {base station (BS)}-to-UAV bistatic links, monostatic sensing provides more direct geometric cues for associating measured MPCs with indoor structures before material comparison. In
BS-to-UAV links, the spatial separation between the transmitter (Tx) and receiver (Rx) makes an observed MPC compatible with multiple candidate surfaces, so the surface responsible for the corresponding material evidence is not directly determined by the sensing geometry and must be inferred.
This attribution ambiguity is amplified in outdoor low-altitude multi-building scenarios, where MPCs from multiple facades are jointly observed at the Rx. The inference is further complicated by rough outdoor building surfaces that induce mixed specular and diffuse propagation. Diffuse components can bias geometry recovery when interpreted under ideal specular assumptions \cite{zhai2025multipath_slam_extended_object},\cite{wielandner2024mimo_multipath_slam}, yet they retain surface-dependent scattering cues that are informative for material inference. Moreover, as the UAV moves, MPC observations, including delay, angle, and received power, are acquired sequentially under time-varying observation geometries. Together, these factors lead to a threefold challenge: probabilistically attributing each incoming MPC to its generating surface under bistatic ambiguity, robustly updating surface-level geometry in the presence of non-ideal specular-diffuse propagation, and recursively refining material posteriors by fusing sequential and uncertain MPC evidence.
\vspace{-0.18\baselineskip}

\begin{figure*}[!t]
\centering
\includegraphics[width=0.98\textwidth]{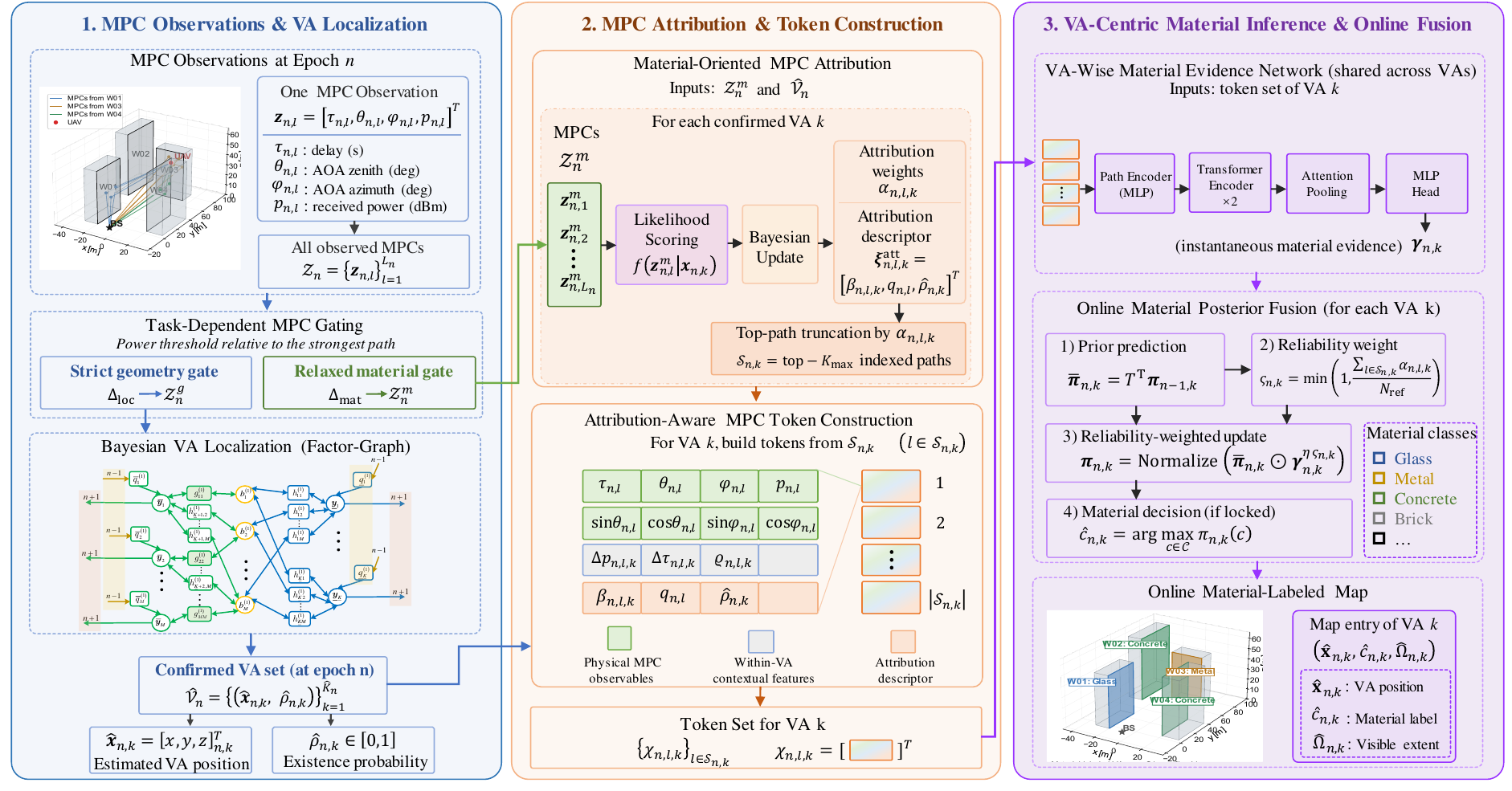}
\caption{Overall framework of the proposed material-labeled environment reconstruction pipeline. The system collects MPC observations, filters candidate paths, performs Bayesian factor-graph VA localization, constructs attribution-aware path tokens using Bayesian multipath attribution, infers material evidence with a transformer-based network, and accumulates online material posteriors to obtain a material-labeled map.}
\label{fig:framework}
\end{figure*}

{We propose an online framework coupling Bayesian VA-based multipath
attribution with material inference, where each VA mirrors the BS across
a reflecting facade (Fig.~\ref{fig:framework}). Compared with the
specular-only {belief propagation-based (BP)}-SLAM in
\cite{leitinger2019belief_propagation_slam}, the proposed framework
explicitly accounts for diffuse facade returns.
Refs.~\cite{liu2026fusion_monostatic_bistatic_isac,zhai2025multipath_slam_extended_object,wielandner2024mimo_multipath_slam} further consider non-ideal or diffuse surface interactions, but do not infer facade materials. Building on these works, we introduce a tail-aware
specular--diffuse likelihood that better accommodates diffuse deviations
while retaining a concentrated specular component for geometric
discrimination, thereby improving VA localization under non-ideal surface
propagation. More importantly, none of these works addresses facade-level
material inference. We further integrate soft MPC-to-VA attribution,
attribution-aware tokens, and recursive neural-evidence fusion to construct
online material-labeled maps.}

{\emph{Notation:}}
Scalars are italic, e.g., $d,p,r$, whereas vectors and matrices are bold,
e.g., $\boldsymbol{x},\boldsymbol{z},\boldsymbol{T}$. The operators
$\Pr(\cdot)$ and $\mathbb{E}[\cdot]$ denote probability and expectation.
Calligraphic letters, e.g., $\mathcal{Z}$ and $\mathcal{C}$, denote sets.
The superscript $(\cdot)^{\mathrm T}$ denotes transpose.
Unless explicitly stated as random, symbols denote deterministic parameters,
indices, or realized observations.

\section{System Model}

This section defines the BS--UAV ISAC setting, the MPC observation model, and the statistical models used by the Bayesian front end.
\vspace{-0.25\baselineskip}
\subsection{Scenario Description and Objective}
\label{subsec:scenario_objective}

We consider a low-altitude BS–UAV ISAC link, where a
moving UAV transmits uplink signals and a fixed antenna array BS receives them for both data communication and radio sensing. The transmitted signals may include uplink reference signals reused for sensing, as well as dedicated  ISAC waveforms. The UAV position $\boldsymbol{u}_n$ at time $n$ is assumed to be available from onboard navigation systems.

{At epoch $n$, the online map is $\mathcal{R}_{n}^{\mathrm{online}}=\{(\hat{\boldsymbol{x}}_{n,k},\hat{c}_{n,k},\hat{\Omega}_{n,k})\}_{k=1}^{\hat{K}_n}$, where $\hat{K}_n$ counts confirmed VAs. Each entry contains a VA position $\hat{\boldsymbol{x}}_{n,k}$, a facade material label $\hat{c}_{n,k}\in\mathcal{C}$ (glass, metal, concrete, or brick), and a coarse visible extent $\hat{\Omega}_{n,k}$. Incoming MPCs update existing entries; new structures enter only after confirmation. Under specular geometry, the VA and known BS positions determine the reflecting plane, while attributed MPCs retain non-ideal propagation effects for material inference. The non-mesh extent is accumulated from visibility and spatial support of attributed MPCs. The final map retains stably confirmed entries after the trajectory.}

\subsection{MPC Observation Model}
\label{subsec:mpc_observation}

\begin{figure*}[!b]
\edef\savedEquationNumber{\number\value{equation}}
\setcounter{equation}{1}
\centering
\vspace{-4.3mm}
\noindent\hrulefill
\vspace{-3pt}
\begin{equation}
\begin{aligned}
& f\big(\boldsymbol{y}_{0:n},\boldsymbol{b}_{1:n},\boldsymbol{m}_{1:n}
 \mid \boldsymbol{z}^{\mathrm{g}}_{1:n};\boldsymbol{u}_{0:n}\big) \\
& \propto
 \prod_{k=1}^{K_{0}} f(\boldsymbol{y}_{0,k};\boldsymbol{u}_{0})
 \prod_{n'=1}^{n}\prod_{l=1}^{m_{n'}} f_{\mathrm{fa}}(\boldsymbol{z}^{\mathrm{g}}_{n',l})
 \prod_{n'=1}^{n}
 \biggl(
 \prod_{k=1}^{\underline{K}_{n'}}
 \underline{q}(\underline{\boldsymbol{x}}_{n',k},\underline{r}_{n',k}\mid
 \boldsymbol{x}_{n'-1,k},r_{n'-1,k})
 \prod_{l=1}^{m_{n'}}
 h_{k}(\underline{\boldsymbol{x}}_{n',k},\underline{r}_{n',k},b_{n',l};
 \boldsymbol{z}^{\mathrm{g}}_{n',l},\boldsymbol{u}_{n'})
\\
& \times
 \prod_{k'=1}^{m_{n'}}
 \bar{q}(\overline{\boldsymbol{x}}_{n',k'},\overline{r}_{n',k'};\boldsymbol{u}_{n'})
 g_{\underline{K}_{n'}+k'}(\overline{\boldsymbol{x}}_{n',k'},\overline{r}_{n',k'},b_{n',k'};
 \boldsymbol{z}^{\mathrm{g}}_{n',k'},\boldsymbol{u}_{n'})
 \prod_{l=k'+1}^{m_{n'}}
 h_{\underline{K}_{n'}+k'}(\overline{\boldsymbol{x}}_{n',k'},\overline{r}_{n',k'},b_{n',l};
 \boldsymbol{z}^{\mathrm{g}}_{n',l},\boldsymbol{u}_{n'})
 \biggr).
\label{eq:F2_bi_only_single_bs}
\end{aligned}
\end{equation}
\setcounter{equation}{\savedEquationNumber}
\noindent\hrulefill
\end{figure*}

To construct the VA-centric material-labeled map defined above, the BS first extracts path-level MPC observations from the received uplink waveforms. Each retained MPC is parameterized by its delay, angle-of-arrival, and power. 
The $l$-th MPC at sensing epoch $n$ is represented as
$\boldsymbol{z}_{n,l}=[\tau_{n,l},\theta_{n,l},\phi_{n,l},p_{n,l}]^{\mathrm T}$,
where $\tau_{n,l}$ is the time of arrival, {$\theta_{n,l}$ and $\phi_{n,l}$ denote the BS-side zenith and azimuth angles of arrival, respectively,} and $p_{n,l}$ is the received power. The geometric component of this observation is denoted by
$\boldsymbol{z}^{\mathrm g}_{n,l}=[d_{n,l},\theta_{n,l},\phi_{n,l}]^{\mathrm T}$,
where $d_{n,l}=c_0\tau_{n,l}$ and $c_0$ is the speed of light. The complete MPC set extracted at epoch $n$ is
$\mathcal{Z}_{n}=\{\boldsymbol{z}_{n,l}\}_{l=1}^{L_n}$,
where $L_n$ denotes the number of extracted MPCs. The geometry-oriented path set used for VA localization  is
$\mathcal{Z}^{\mathrm g}_{n}=\{\boldsymbol{z}^{\mathrm g}_{n,l}\}_{l\in\mathcal{L}^{\mathrm g}_n}$,
where $\mathcal{L}^{\mathrm g}_n\subseteq\{1,\ldots,L_n\}$ denotes the indices retained by the geometry-oriented gate. Similarly, the material-oriented path set used for material evidence extraction is
$\mathcal{Z}^{\mathrm m}_{n}=\{\boldsymbol{z}_{n,l}\}_{l\in\mathcal{L}^{\mathrm m}_n}$,
where $\mathcal{L}^{\mathrm m}_n\subseteq\{1,\ldots,L_n\}$ denotes the indices retained by the material-oriented gate. The two index sets need not coincide: the geometry-oriented gate is stricter to suppress diffuse components that may impair VA localization, whereas
the material-oriented gate retains more diffuse and weak reflected MPCs carrying material cues.

\subsection{Statistical Models}
\label{subsec:statistical_models}

{The following VA state and likelihood models support VA localization from $\mathcal{Z}^{\mathrm g}_{n}$ and subsequent attribution of $\mathcal{Z}^{\mathrm m}_{n}$ to the confirmed VAs.}

\emph{VA state model:}
We follow the bistatic VA formulation in \cite{liu2026fusion_monostatic_bistatic_isac}, omitting monostatic sensing variables and the BS index. The $k$-th candidate facade is represented by the augmented VA state $\boldsymbol{y}_{n,k}=[\boldsymbol{x}_{n,k}^{\mathrm T},r_{n,k}]^{\mathrm T}$, where $\boldsymbol{x}_{n,k}\in\mathbb{R}^{3}$ is the latent VA position and $r_{n,k}\in\{0,1\}$ is its existence indicator. Thus, each facade-level VA is treated as a random entity whose position and existence are inferred from the posterior distribution conditioned on $\mathcal{Z}^{\mathrm g}_{n}$. Since facades are static, the VA position follows a nearly static transition model, e.g., $\boldsymbol{x}_{n,k}=\boldsymbol{x}_{n-1,k}+\boldsymbol{w}_{n,k}$ with small process noise $\boldsymbol{w}_{n,k}$. The existence variable follows a survival-birth mechanism for propagated and newly initialized
VA hypotheses \cite{liu2026fusion_monostatic_bistatic_isac}.

\emph{Bistatic measurement and likelihood model:}
\begingroup
Let $\boldsymbol{p}_{\mathrm B}$ denote the {known, fixed} BS position. Given the UAV position
$\boldsymbol{u}_{n}$ and a candidate VA position $\boldsymbol{x}_{n,k}$, we use
$f(\boldsymbol{z}^{\mathrm g}_{n,l}\mid \boldsymbol{x}_{n,k};\boldsymbol{u}_{n},\boldsymbol{p}_{\mathrm B})$
to denote the bistatic delay-angle likelihood model. It evaluates the geometric
consistency between the $l$-th MPC and the $k$-th VA using the reflected path
length and the BS-side AoA. Since the VA is the mirror image of the BS with
respect to the reflecting facade, the corresponding reflection plane has normal
direction $\boldsymbol{n}_{n,k}=\frac{\boldsymbol{x}_{n,k}-\boldsymbol{p}_{\mathrm B}}{\|\boldsymbol{x}_{n,k}-\boldsymbol{p}_{\mathrm B}\|}$
and passes through the midpoint $\boldsymbol{p}^{\mathrm{mid}}_{n,k}=\frac{\boldsymbol{x}_{n,k}+\boldsymbol{p}_{\mathrm B}}{2}$.
The specular reflection point is obtained by intersecting the UAV--VA line with
this plane: $\boldsymbol{s}_{n,k}=\boldsymbol{u}_{n}+\alpha_{n,k}(\boldsymbol{x}_{n,k}-\boldsymbol{u}_{n})$,
where $\alpha_{n,k}=\frac{\boldsymbol{n}_{n,k}^{\mathrm T}(\boldsymbol{p}^{\mathrm{mid}}_{n,k}-\boldsymbol{u}_{n})}{\boldsymbol{n}_{n,k}^{\mathrm T}(\boldsymbol{x}_{n,k}-\boldsymbol{u}_{n})}$.
The predicted path length is $\hat d_{n,k}=\|\boldsymbol{x}_{n,k}-\boldsymbol{u}_{n}\|$.
With $\boldsymbol{v}_{n,k}=\boldsymbol{s}_{n,k}-\boldsymbol{p}_{\mathrm B}=[v_{n,k,x},v_{n,k,y},v_{n,k,z}]^{\mathrm T}$,
the predicted BS-side angles are $\hat\theta_{n,k}=\arccos(v_{n,k,z}/\|\boldsymbol{v}_{n,k}\|)$
and $\hat\phi_{n,k}=\operatorname{atan2}(v_{n,k,y},v_{n,k,x})$.
For the observed MPC $\boldsymbol{z}^{\mathrm g}_{n,l}$, the
path-length, zenith-angle, and azimuth-angle residuals are defined as
$\Delta d_{n,l,k}=d_{n,l}-\hat d_{n,k}$,
$\Delta\theta_{n,l,k}=\theta_{n,l}-\hat\theta_{n,k}$, and
$\Delta\phi_{n,l,k}=\operatorname{wrap}(\phi_{n,l}-\hat\phi_{n,k})$, respectively,
where $\operatorname{wrap}(\cdot)$ maps the azimuth residual to $[-\pi,\pi)$.
\endgroup
For rough building facades, an observed MPC may be a
dominant specular reflection or a diffuse component around the same facade.
To retain geometric discrimination
while tolerating diffuse deviations, we
use a tail-aware specular-diffuse likelihood
\begin{equation}
\begin{aligned}
&{f(\boldsymbol{z}^{\mathrm g}_{n,l}\mid \boldsymbol{x}_{n,k};\boldsymbol{u}_{n},\boldsymbol{p}_{\mathrm B})} \\
&\quad =
{\frac{
f_{\mathrm{sp}}(\boldsymbol{z}^{\mathrm g}_{n,l}\mid \boldsymbol{x}_{n,k};\boldsymbol{u}_{n},\boldsymbol{p}_{\mathrm B})
+\lambda_{\mathrm d}
f_{\mathrm{diff}}(\boldsymbol{z}^{\mathrm g}_{n,l}\mid \boldsymbol{x}_{n,k};\boldsymbol{u}_{n},\boldsymbol{p}_{\mathrm B})
}{1+\lambda_{\mathrm d}}},
\end{aligned}
\label{eq:tail_likelihood_secII}
\end{equation}
where $f_{\mathrm{sp}}(\cdot)$ is {a probability density} concentrated around the ideal bistatic VA
geometry, and $f_{\mathrm{diff}}(\cdot)$ is {a broader probability density accounting for diffuse deviations.}
{The parameter $\lambda_{\mathrm d}\geq 0$ controls the relative diffuse contribution, and the factor $(1+\lambda_{\mathrm d})^{-1}$ ensures that the mixture is normalized.}

\emph{Two-stage use:}
During Bayesian VA localization, 
\eqref{eq:tail_likelihood_secII} is combined with the localization-stage data
association variable over $\mathcal{Z}^{\mathrm g}_{n}$ to update VA states. After VA confirmation, the same likelihood principle is reapplied to
$\mathcal{Z}^{\mathrm m}_{n}$ to compute material-stage MPC-to-VA
attribution probabilities.

\section{Bayesian VA Localization and Multipath Attribution}
\label{sec:bayesian_va_attribution}

This section describes the Bayesian front end that first uses $\mathcal{Z}^{\mathrm g}_{n}$ to infer VA states and existence probabilities through a factor
graph, then re-scores $\mathcal{Z}^{\mathrm m}_{n}$ against confirmed VAs to obtain material-oriented attribution weights.

\stepcounter{equation}
\subsection{Bayesian VA Posterior Factorization}
\label{subsec:va_posterior}

For Bayesian VA localization, only $\mathcal{Z}^{\mathrm g}_{n}$ is used.
Locally re-indexing the retained MPCs gives
$\mathcal{Z}^{\mathrm g}_{n}=\{\boldsymbol{z}^{\mathrm g}_{n,l}\}_{l=1}^{m_n}$,
where $m_n=|\mathcal{L}^{\mathrm g}_n|$ is the number of geometry-oriented
MPCs retained at epoch $n$. Let
$\boldsymbol{y}_{n,k}=[\boldsymbol{x}_{n,k}^{\mathrm T},r_{n,k}]^{\mathrm T}$
denote the augmented state of the $k$-th VA, and let
$\boldsymbol{b}_{n}=\{b_{n,l}\}_{l=1}^{m_n}$ denote the localization-stage
data-association vector, where $b_{n,l}=k$ indicates that the $l$-th
geometry-oriented MPC is associated with the $k$-th VA, while $b_{n,l}=0$
denotes an unassigned or clutter hypothesis.

The posterior factorization follows the bistatic factor-graph model in
\cite{liu2026fusion_monostatic_bistatic_isac}. In particular, the prior,
false-alarm, transition, birth, and measurement-association factors retain the
same forms. {For brevity, the dependence on the known, fixed BS position $\boldsymbol{p}_{\mathrm B}$ is suppressed in \eqref{eq:F2_bi_only_single_bs}.} The resulting posterior is written as (\ref{eq:F2_bi_only_single_bs}).

In \eqref{eq:F2_bi_only_single_bs}, $f_{\mathrm{fa}}(\cdot)$ is the
false-alarm density, $\underline{q}(\cdot)$ and $\bar{q}(\cdot)$ are
legacy-state transition and birth densities, and $h_k(\cdot)$
and $g_k(\cdot)$ are measurement-association factors. The task-specific component is the bistatic delay-angle likelihood 
$
    \ell^{\mathrm g}_{n,l,k}
    =
    {f(\boldsymbol{z}^{\mathrm g}_{n,l}\mid\boldsymbol{x}_{n,k};\boldsymbol{u}_{n},\boldsymbol{p}_{\mathrm B})},
    \label{eq:bistatic_likelihood}
$
where $f(\cdot)$ follows 
\eqref{eq:tail_likelihood_secII}. Its two components are
realized in the delay-angle domain as
	${f_{\nu}(\boldsymbol{z}^{\mathrm g}_{n,l}\mid\boldsymbol{x}_{n,k};\boldsymbol{u}_{n},\boldsymbol{p}_{\mathrm B})}
	=\allowbreak
	\mathcal{N}(\Delta d_{n,l,k};0,\sigma^{2}_{d,\nu})
\allowbreak
\mathcal{N}_{\mathrm w}(\Delta\theta_{n,l,k};0,\sigma^{2}_{\theta,\nu})
\allowbreak
\mathcal{N}_{\mathrm w}(\Delta\phi_{n,l,k};0,\sigma^{2}_{\phi,\nu})$,
where $\nu\in\{\mathrm{sp},\mathrm{diff}\}$, and
$\Delta d_{n,l,k}$, $\Delta\theta_{n,l,k}$, and $\Delta\phi_{n,l,k}$ are the
{path-length, zenith-angle, and azimuth-angle residuals with respect to the
BS-side reflection geometry}. The angular residuals are modeled by wrapped
Gaussian densities $\mathcal{N}_{\mathrm w}(\cdot)$, with
$\sigma^{2}_{\cdot,\mathrm{diff}}>\sigma^{2}_{\cdot,\mathrm{sp}}$.
The posterior yields the VA existence probability
$\hat{\rho}_{n,k}=\Pr(r_{n,k}=1\mid\mathcal{Z}^{\mathrm g}_{1:n}{;\boldsymbol{u}_{0:n},\boldsymbol{p}_{\mathrm B}})$ and the
conditional MMSE estimate
$\hat{\boldsymbol{x}}_{n,k}=
\mathbb{E}[\boldsymbol{x}_{n,k}\mid r_{n,k}=1,\mathcal{Z}^{\mathrm g}_{1:n}{;\boldsymbol{u}_{0:n},\boldsymbol{p}_{\mathrm B}}]$,
following \cite{liu2026fusion_monostatic_bistatic_isac}. VAs with
$\hat{\rho}_{n,k}$ above a threshold form the confirmed VA set
for material attribution.

\subsection{Material-Oriented MPC Attribution}
\label{subsec:material_attribution}

The localization-stage association vector $\boldsymbol{b}_{n}$ is not passed
directly to the material-recognition network. After VA confirmation, the material stage re-scores the richer material-oriented path set $\mathcal{Z}^{\mathrm m}_{n}$ against the confirmed VA entities.

Let
$\widehat{\mathcal{V}}_{n}
=\{(\hat{\boldsymbol{x}}_{n,k},\hat{\rho}_{n,k})\}_{k=1}^{\hat{K}_n}$
denote the confirmed VA set, where $\hat{\boldsymbol{x}}_{n,k}$ is the
estimated position of the $k$-th VA and $\hat{\rho}_{n,k}$ is its 
existence reliability. For a material-stage MPC $\boldsymbol{z}^{\mathrm m}_{n,l}$,
let $\boldsymbol{z}^{\mathrm{m,g}}_{n,l}$ be its geometric component.  
Conditioned on $\widehat{\mathcal{V}}_{n}$, each path--VA edge is re-scored
using the same tail-aware bistatic likelihood defined in
Sec.~\ref{subsec:statistical_models}. Specifically, we define
\begin{equation}
    \ell^{\mathrm m}_{n,l,k}
    =
    f(\boldsymbol{z}^{\mathrm{m,g}}_{n,l}
    \mid
    \hat{\boldsymbol{x}}_{n,k};\boldsymbol{u}_{n}{,\boldsymbol{p}_{\mathrm B}}),
    \qquad
    s_{n,l,k}
    =
    \kappa\hat{\rho}_{n,k}\ell^{\mathrm m}_{n,l,k},
    \label{eq:material_score}
\end{equation}
where $\kappa$ is the likelihood-ratio scale and the background score is
normalized to one. Let $S_{n,l}=\sum_{j=1}^{\hat{K}_n}s_{n,l,j}$. For
$S_{n,l}>0$, the material-stage attribution quantities are defined as
\begin{equation}
\setlength{\arraycolsep}{2pt}
\begin{gathered}
    q_{n,l}=\frac{S_{n,l}}{1+S_{n,l}},\qquad
    \beta_{n,l,k}=\frac{s_{n,l,k}}{S_{n,l}},\\[-0.5mm]
    \alpha_{n,l,k}=q_{n,l}\beta_{n,l,k}
    =
    \frac{s_{n,l,k}}{1+S_{n,l}} .
\end{gathered}
\label{eq:q_beta}
\end{equation}
If $S_{n,l}=0$, the path is left unassigned by setting $q_{n,l}=0$ and
$\alpha_{n,l,k}=0$. Here, $q_{n,l}$ is the confirmed-VA confidence, measuring
whether the path is reliably explained by the confirmed VA set. The conditional
weight $\beta_{n,l,k}$ describes how the path is allocated among confirmed VAs
once it is considered VA-related. Their product $\alpha_{n,l,k}$ is the
unconditional VA-specific attribution, which measures the probability that the
$l$-th MPC is attributed to the $k$-th confirmed VA.

This decomposition is used in the material-token construction. For each
MPC--VA pair, we define the attribution descriptor
$\boldsymbol{\xi}^{\mathrm{att}}_{n,l,k}
=[\beta_{n,l,k},q_{n,l},\hat{\rho}_{n,k}]^{\mathrm T}$,
which summarizes the conditional VA allocation, the path-level confirmed-VA
confidence, and the VA existence reliability. The scalar
$\alpha_{n,l,k}$ is used as the attribution weight for
ranking or selecting material-stage paths associated with VA $k$. The selected paths are tokenized using their physical MPC observables together
with $\boldsymbol{\xi}^{\mathrm{att}}_{n,l,k}$, enabling facade-specific material
evidence extraction from attributed MPCs.

\section{Attribution-Aware VA-Centric Material Inference}
\label{sec:material_inference}
 For each confirmed VA, the
material module constructs a VA-centric set of attributed MPC tokens, extracts
instantaneous material evidence with a shared neural network, and recursively
fuses it into an online material posterior.

\subsection{Attribution-Aware MPC Token Construction}
\label{subsec:mpc_token}

For the $k$-th confirmed VA at epoch $n$, we retain material-stage MPCs with
positive conditional allocation $\beta_{n,l,k}>0$ and sort them by
$\alpha_{n,l,k}=q_{n,l}\beta_{n,l,k}$ in \eqref{eq:q_beta}. This ordering is used
only for deterministic top-path truncation and for the rank feature below; it is
not an additional physical assumption. Let $\mathcal{S}_{n,k}$ denote the
resulting VA-centric path index set after truncation. The normalized rank is
$
    \varrho_{n,l,k}
    =
    \frac{\operatorname{rank}_{n,l,k}}
    {\max(|\mathcal{S}_{n,k}|-1,1)} ,
$
where $\operatorname{rank}_{n,l,k}=0$ corresponds to the highest-attribution
retained path. Thus $\varrho_{n,l,k}$ records the ordinal attribution priority
already used by the implementation to keep the most reliable MPC evidence.
For each selected path $l\in\mathcal{S}_{n,k}$, we define
$\Delta p_{n,l,k}=p_{n,l}-p^{\max}_{n,k}$ and
$\Delta\tau_{n,l,k}=\tau_{n,l}-\tau^{\min}_{n,k}$, where
$p^{\max}_{n,k}=\max_{l'\in\mathcal{S}_{n,k}}p_{n,l'}$ and
$\tau^{\min}_{n,k}=\min_{l'\in\mathcal{S}_{n,k}}\tau_{n,l'}$. Let
{$\boldsymbol{o}_{n,l}=[p,\tau,\theta,\phi,\sin\theta,\cos\theta,
\sin\phi,\cos\phi]_{n,l}^{\mathrm T}$} collect the physical MPC observables.
The final attribution-aware token for
path $l$ with respect to VA $k$ is
\begin{equation}
    \boldsymbol{\chi}_{n,l,k}
    =
    \big[
    \boldsymbol{o}_{n,l}^{\mathrm T},
    \Delta p_{n,l,k},
    \Delta \tau_{n,l,k},
    \varrho_{n,l,k},
    (\boldsymbol{\xi}^{\mathrm{att}}_{n,l,k})^{\mathrm T}
    \big]^{\mathrm T},
    \label{eq:material_token}
\end{equation}
Thus, each token contains both physical propagation features and probabilistic
path-to-VA attribution information.

\subsection{VA-Wise Material Evidence Extraction}
\label{subsec:va_material_network}

The material evidence network is shared across all confirmed VAs and is applied
independently to each VA-centric token set, as illustrated in
Fig.~\ref{fig:framework}. For VA $k$, the token encoder maps each attributed MPC
token to $\boldsymbol{h}_{n,l,k}=f_{\psi}(\boldsymbol{\chi}_{n,l,k})$; a
padding-masked transformer then produces
$\widetilde{\boldsymbol{H}}_{n,k}=\operatorname{Transformer}
(\{\boldsymbol{h}_{n,l,k}\}_{l\in\mathcal{S}_{n,k}})$ to model interactions
among paths assigned to the same VA. A learned attention-pooling head summarizes
these path-level embeddings as
$\boldsymbol{g}_{n,k}=\operatorname{AttnPool}(\widetilde{\boldsymbol{H}}_{n,k})$,
which is fed to a classifier to obtain the instantaneous material evidence
\begin{equation}
    \boldsymbol{\gamma}_{n,k}
    =
    \operatorname{softmax}
    \big(f_{\omega}(\boldsymbol{g}_{n,k})\big),
    \label{eq:instant_material_evidence}
\end{equation}
where the $c$-th entry of $\boldsymbol{\gamma}_{n,k}$ represents the
instantaneous probability that the facade associated with VA $k$ belongs to
material class $c\in\mathcal{C}$.

\subsection{Online Material Posterior Fusion}
\label{subsec:online_material_update}

Because both MPC visibility and attribution quality vary along the UAV
trajectory, the instantaneous material evidence is fused recursively at the VA
level. Let $\boldsymbol{\pi}_{n,k}$ denote the accumulated material posterior of
VA $k$ after epoch $n$. Before incorporating the current network output
$\boldsymbol{\gamma}_{n,k}$, we first predict the prior material belief and
compute the reliability of the current attributed MPC evidence as
\begin{align}
    \bar{\boldsymbol{\pi}}_{n,k} 
    &=
    \boldsymbol{T}^{\mathrm T}\boldsymbol{\pi}_{n-1,k},
    \label{eq:material_prediction}\\
    \zeta_{n,k}
    &=
    \min\left(1,
    \frac{\sum_{l\in\mathcal{S}_{n,k}}\alpha_{n,l,k}}{N_{\mathrm{ref}}}
    \right),
    \label{eq:evidence_reliability}
\end{align}
where $\boldsymbol{T}$ is close to diagonal because facade materials are static
over the trajectory. The numerator in \eqref{eq:evidence_reliability} measures
the effective amount of attributed MPC evidence for VA $k$, while
$N_{\mathrm{ref}}$ is the reference support required for a fully reliable
update. Hence, $\zeta_{n,k}\in[0,1]$ acts as an adaptive reliability gate for
the current network output.
The posterior is updated by
\begin{equation}
    \boldsymbol{\pi}_{n,k}
    =
    \operatorname{Normalize}
    \left(
    \bar{\boldsymbol{\pi}}_{n,k}
    \odot
    \boldsymbol{\gamma}_{n,k}^{\,\eta\zeta_{n,k}}
    \right),
    \label{eq:material_posterior_update}
\end{equation}
{where $\odot$ denotes element-wise multiplication and $\eta$ controls evidence strength. For each material
class $c$, $\pi_{n,k}(c)\propto\bar{\pi}_{n,k}(c)\gamma_{n,k}(c)^{\eta\zeta_{n,k}}$. Weak attributed support reduces $\zeta_{n,k}$ and tempers the network evidence, whereas sufficient support permits a stronger update. After the trajectory, $\hat{c}_{k}=\arg\max_{c\in\mathcal{C}}\pi_{N,k}(c)$. This label, the confirmed VA position, and the visible extent form the material-labeled map entry.}

\section{Simulation Setup and Evaluation Protocol}

\begin{figure}[!htbp]
\centering
\includegraphics[width=\columnwidth]{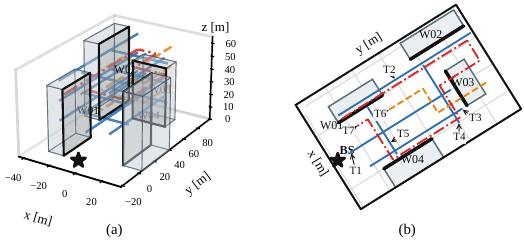}
\caption{Representative urban street scenario and UAV trajectories. Left: three-dimensional view. Right: top-view trajectory layout.}
\label{fig:s1_scene}
\end{figure}

We evaluate the proposed framework in the representative urban street scenario
shown in Fig.~\ref{fig:s1_scene}. A fixed BS is located at $(0,-18,8)^{\mathrm T}$~m, and
the UAV moves through the low-altitude street corridor. The evaluated facades
are W01--W04.

The MPC observations are generated with Sionna RT at 28~GHz using a transmit power
of 44~dBm.  The target facades are drawn from the material set
$\mathcal{C}=\{\mathrm{glass},\mathrm{metal},\mathrm{concrete},\mathrm{brick}\}$,
and each facade is modeled by the corresponding Sionna ITU radio material
configured according to ITU-R P.2040-4 \cite{itu2025p2040}. Across all
materials, we use a thickness of 0.1~m, directive scattering, a
cross-polarization coefficient of 0.4, and $\alpha_r=4$; the
material-dependent scattering coefficients are listed in
Table~\ref{tab:material_rt_config}.

\begin{table}[!t]
\caption{{Material-dependent scattering coefficients used in Sionna RT}}
\label{tab:material_rt_config}
\centering
\begin{tabular}{lc|lc}
\hline
Material & Scattering coeff. & Material & Scattering coeff. \\
\hline
Glass & 0 & Metal & 0.15 \\
Concrete & 0.273 & Brick & 0.4 \\
\hline
\end{tabular}
\end{table}

To obtain measurement-level MPC observations, RT path length, {BS-side zenith/azimuth AoA}, and received
power are perturbed by zero-mean Gaussian noise with standard deviations
$\sigma_d=0.2$~m, $\sigma_\theta=\sigma_\phi=0.25^\circ$, and
$\sigma_p=1$~dB. Noisy paths are retained by a snapshot-wise dynamic-range rule:
$\Delta_{\mathrm{loc}}=20$~dB for VA localization and
$\Delta_{\mathrm{mat}}=50$~dB for material attribution, both relative to the
strongest path in the snapshot. 

{The above perturbations capture random uncertainty in the extracted MPC parameters, including imperfect channel estimation and receiver noise. {Systematic GPS, synchronization, and calibration errors are not explicitly modeled in the present evaluation. A natural extension is to augment} the factor graph with low-dimensional bias states, such as UAV position and common delay/angle offsets, which can be jointly inferred with the VA states when sufficiently observable.}

We use 48 balanced four-facade material configurations, split into 32/8/8 for
training, validation, and testing, with disjoint material vectors
$[c_{\mathrm{W01}},c_{\mathrm{W02}},c_{\mathrm{W03}},c_{\mathrm{W04}}]$ across
splits. The trajectory split is also disjoint: T1--T5 for training, T6 for
validation, and the held-out T7 trajectory for testing. Each trajectory is
sampled at approximately 1~m intervals. Training and validation use multiple
UAV heights, while the final test uses T7 at the unseen height $h=45$~m.

Each material-recognition sample is the VA-centric token set
$\{\boldsymbol{\chi}_{n,l,k}\}_{l\in\mathcal{S}_{n,k}}$ in
\eqref{eq:material_token}, sorted by $\alpha_{n,l,k}$. The resulting dataset contains 111,803 training, 8,526 validation, and
4,241 test VA-wise samples. For online evaluation, $\boldsymbol{\pi}_{0,k}$ is
uniform, the transition matrix has diagonal probability 0.995 and off-diagonal
probability $0.005/3$. {The validation search uses $\eta\in\{0.1,0.2,0.3,0.5\}$ and $N_{\mathrm{ref}}\in\{4,8,16\}$, prioritizing final-VA macro-F1 and then accuracy; remaining ties favor smaller values. Larger $\eta$ strengthens each evidence update, whereas larger $N_{\mathrm{ref}}$ requires more attributed support. The two dynamic-range gates balance localization reliability and material-evidence coverage.} A final material decision is locked when the top posterior
exceeds 0.9 for four consecutive reliable observations; otherwise, the last
reliable posterior is used.

\section{Results and Discussion}
\begin{figure}[!t]
\centering
\includegraphics[width=\columnwidth]{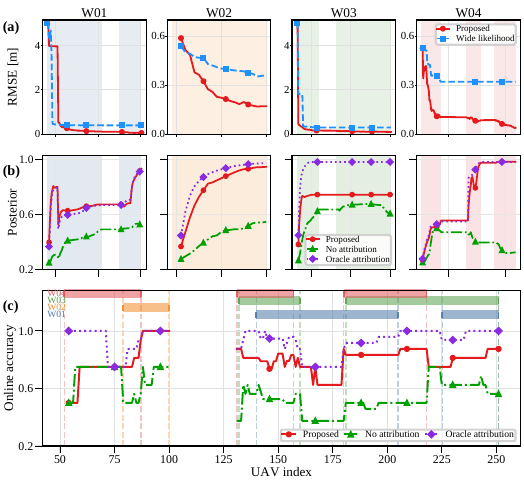}
\caption{Combined online evaluation for the street scenario. 
(a) Per-facade VA localization RMSE for the proposed method and the wide-likelihood baseline 
\cite{liu2026fusion_monostatic_bistatic_isac}, \cite{wielandner2024mimo_multipath_slam}. 
(b) Online posterior assigned to the correct material class for each facade. 
(c) Online material-recognition accuracy along the held-out T7 trajectory, where colored dashed bars at the top indicate the visible evaluation intervals of W01--W04.}
\label{fig:combined_results}
\vspace{0.5em}
\includegraphics[width=1\columnwidth]{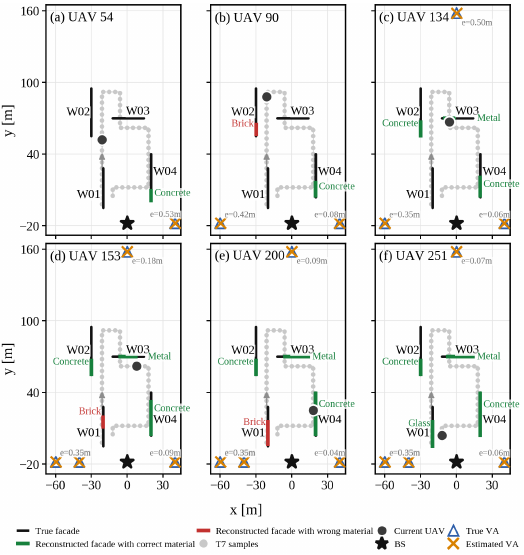}
\caption{Online evolution of the material-labeled reconstruction on the held-out T7 trajectory for one representative material configuration from the test set.}
\label{fig:online_reconstruction_evolution}
\end{figure}
We evaluate the proposed framework on the held-out T7 trajectory in terms of VA
localization, online material posterior evolution, and final material-labeled
map accuracy. Fig.~\ref{fig:combined_results} reports the online metrics, while
Fig.~\ref{fig:online_reconstruction_evolution} shows representative map updates.
For material recognition, all compared methods use the same network architecture
and posterior fusion rule, but differ in MPC-to-facade attribution. The
no-attribution baseline uses mixed MPC evidence, whereas oracle attribution
uses ground-truth surface labels as an upper bound.
{Existing non-ideal-surface SLAM methods do not directly target facade-level material labeling. Therefore, we conduct controlled comparisons under a shared material-inference pipeline to isolate the contribution of MPC-to-facade attribution.}

Fig.~\ref{fig:combined_results}(a) shows that the proposed tail-aware likelihood
achieves stable VA localization across the visible facades. Compared with the
wide-likelihood baseline, it provides faster convergence and lower residual
errors, indicating that the specular component preserves VA separability while
the diffuse tail improves robustness to non-ideal facade scattering.
Fig.~\ref{fig:combined_results}(b) shows that the proposed material posterior
closely follows the oracle trend on most facades and remains clearly above the
no-attribution baseline.{ W03} is more challenging mainly because it is observed over a shorter effective interval along T7, providing less reliable material evidence for posterior accumulation. Consequently, its material posterior converges more slowly and is more sensitive to attribution uncertainty, whereas the oracle directly uses ground-truth facade associations. The online accuracy in Fig.~\ref{fig:combined_results}(c)
shows the same map-level trend: the proposed method stays well above the
no-attribution baseline and remains close to oracle attribution over the
visible evaluation intervals.

Fig.~\ref{fig:online_reconstruction_evolution} illustrates the online
reconstruction process. As more VA-attributed MPC evidence is accumulated, the
material-labeled map is progressively completed and corrected. The stable VA
locations indicate that the remaining material errors are mainly caused by
difficult attribution or weak material evidence, rather than large geometric
drift.

\begin{table}[!t]
\caption{Final material-labeled reconstruction performance on the held-out T7 test trajectory.}
\label{tab:final_reconstruction}
\centering
\footnotesize
\begin{tabular}{lcccc}
\hline
Method & RMSE (m) & Acc. & Macro-F1 & Correct \\
\hline
Proposed & 0.0807 & 0.9375 & 0.9365 & 30/32 \\
No attribution & 0.0807 & 0.5000 & 0.4507 & 16/32 \\
Oracle attribution & 0.0807 & 1.0000 & 1.0000 & 32/32 \\
\hline
\end{tabular}
\end{table}

Table~\ref{tab:final_reconstruction} summarizes the final reconstruction performance. All material-attribution variants share the same VA localization, yielding an identical geometry RMSE of 0.0807~m. The proposed method correctly labels 30/32 facade instances, achieving 93.75\% accuracy and a macro-F1 of 0.9365, versus 50.00\% for no attribution and 100\% for oracle attribution. These results confirm that probabilistic facade attribution drives the material-recognition gain, while the remaining oracle gap reflects residual attribution uncertainty.
{Factor-graph cost grows with retained MPCs, candidate VAs, and message updates. On a CUDA GPU server, material-network inference and online posterior fusion take about 7.9~ms per receiver snapshot for BP-soft attribution. End-to-end real-time feasibility therefore remains unverified.}
{Denser streets may increase attribution ambiguity among geometrically similar facades, which is naturally represented by the proposed soft Bayesian attribution. Non-standard or weathered materials can be further accommodated by extending the training set with more diverse material responses.}

\FloatBarrier

\section{Conclusion}

This paper developed an online framework for material-labeled reconstruction in low-altitude BS--UAV ISAC, integrating Bayesian VA localization, tail-aware multipath attribution, and VA-centric posterior fusion. Ray-tracing results show that probabilistic attribution enables reliable material inference from mixed sequential MPCs, achieving 93.75\% final facade-level accuracy on a held-out trajectory while maintaining accurate VA localization.
\bibliographystyle{IEEEtran}
\bibliography{references}

@article{jiang2025isac_low_altitude_economy,
  author   = {Jiang, Y. and others},
  title    = {Integrated Sensing and Communication for Low Altitude Economy: Opportunities and Challenges},
  journal  = {IEEE Commun. Mag.},
  volume   = {63},
  number   = {12},
  pages    = {72--78},
  month    = dec,
  year     = {2025},
  doi      = {10.1109/MCOM.001.2400685}
}

@ARTICLE{leitinger2019belief_propagation_slam,
  author  = {Leitinger, E. and others},
  title   = {A Belief Propagation Algorithm for Multipath-Based {SLAM}},
  journal = {IEEE Trans. Wireless Commun.},
  volume  = {18},
  number  = {12},
  pages   = {5613--5629},
  month   = dec,
  year    = {2019}
}

@techreport{itu2025p2040,
  title        = {Effects of building materials and structures on radio-wave propagation in the range of 1 {MHz} to 450 {GHz}},
  author       = {{ITU-R}},
  institution  = {International Telecommunication Union},
  type         = {Recommendation ITU-R P.2040-4},
  year         = {2025},
  month        = sep,

}

@ARTICLE{chang2025environment_reconstruction_thz,
  author  = {Chang, Z. and others},
  title   = {Environment Reconstruction With Multi-Targets Reflectors-Merged Sensing Method Based on {THz} Single-Sided Channel Characteristics},
  journal = {IEEE Wireless Commun. Lett.},
  volume  = {14},
  number  = {5},
  pages   = {1471--1475},
  month   = may,
  year    = {2025}
}

@ARTICLE{song2026dynamic_environment_reconstruction_isac,
  author   = {Song, J. and others},
  title    = {Deep Learning-Based Dynamic Environment Reconstruction for Vehicular {ISAC} Scenarios},
  journal  = {IEEE Trans. Wireless Commun.},
  volume   = {25},
  pages    = {14198--14211},
  year     = {2026},
  doi      = {10.1109/TWC.2026.3676063}
}

@INPROCEEDINGS{yang2025envrecongpt,
  author    = {Yang, T. and others},
  title     = {{EnvReconGPT}: A Generative {AI} Model for Wireless Environment Reconstruction in the {6G} Metaverse},
  booktitle = {IEEE INFOCOM 2025 - IEEE Conference on Computer Communications Workshops (INFOCOM WKSHPS)},
  address   = {London, United Kingdom},
  pages     = {1--6},
  year      = {2025},
  doi       = {10.1109/INFOCOMWKSHPS65812.2025.11152849}
}

@article{liu2026fusion_monostatic_bistatic_isac,
  author = {Liu, M. and others},
  title = {Fusion of Monostatic and Bistatic Sensing for {ISAC}-Enabled Low-Altitude Environment Mapping},
  journal = {IEEE Trans. Commun.},
  volume = {74},
  pages = {12451--12467},
  year = {2026},
  doi = {10.1109/TCOMM.2026.3712538}
}

@ARTICLE{lyu2025hybrid_channel_modeling_thz,
  author   = {Lyu, Y. and others},
  title    = {Hybrid Channel Modeling and Environment Reconstruction for Terahertz Monostatic Sensing},
  journal  = {IEEE Trans. Wireless Commun.},
  volume   = {24},
  number   = {10},
  pages    = {8492--8504},
  month    = oct,
  year     = {2025},
  doi      = {10.1109/TWC.2025.3567292}
}

@ARTICLE{zhai2025multipath_slam_extended_object,
  author  = {Zhai, S. and others},
  title   = {Multipath-Based {SLAM} Exploiting Extended Object Estimation and Classification},
  journal = {IEEE Trans. Wireless Commun.},
  volume  = {24},
  number  = {8},
  pages   = {7029--7045},
  month   = aug,
  year    = {2025}
}

@INPROCEEDINGS{wielandner2024mimo_multipath_slam,
  author    = {Wielandner, L. and others},
  title     = {{MIMO} Multipath-Based {SLAM} for Non-Ideal Reflective Surfaces},
  booktitle = {Proc. 27th Int. Conf. Inf. Fusion (FUSION)},
  address   = {Venice, Italy},
  pages     = {1--8},
  month     = jul,
  year      = {2024}
}

@article{LIU2026130,
title = {Cooperative Sensing for {6G ISAC}: Concept, Key Technologies, Performance Evaluation, and Field Trial},
journal = {Engineering},
volume = {56},
pages = {130-148},
year = {2026},
issn = {2095-8099},
doi = {https://doi.org/10.1016/j.eng.2025.08.033},

author = {Guangyi Liu and others},
}

@article{sun2026modeling_land_to_ship,
  author = {Sun, S. and others},
  title = {Modeling and Analysis of Land-to-Ship Maritime Wireless Channels at 5.8 {GHz}},
  journal = {IEEE Trans. Wireless Commun.},
  volume = {25},
  pages = {10051--10065},
  year = {2026}
}

@article{zhang2026diffuse_scattering_measurements,
  author = {Zhang, T. and others},
  title = {Diffuse Scattering Measurements and Mechanism Analysis at 8, 12, and 28 {GHz} for Typical Building Surfaces},
  journal = {npj Wirel. Technol.},
  volume = {2},
  number = {1},
  month = jan,
  year = {2026}
}

\end{document}